%% file: main.tex
\documentclass[a4paper]{article}

\usepackage{INTERSPEECH2020}
\usepackage{longtable}
\usepackage[colorlinks]{hyperref}       
\usepackage{url}            
\usepackage{booktabs}       
\usepackage{amsfonts}       
\usepackage{nicefrac}       
\usepackage{microtype}      
\usepackage{hyperref}
\usepackage{amsmath,amssymb,amsthm}
\usepackage{verbatim,float,url,enumerate}
\usepackage{graphicx,subfigure,psfrag}
\usepackage{xcolor}
\usepackage{microtype}
\usepackage{graphics}
\usepackage{tikz}
\usepackage{bbm}
\usepackage[english]{babel}
\usepackage[utf8]{inputenc}
\usetikzlibrary{bayesnet}
\usetikzlibrary{arrows}
\usetikzlibrary{backgrounds,positioning}
\usetikzlibrary{decorations.pathreplacing}
\usepackage{caption}
\usepackage{mathtools}
\usepackage{wrapfig,lipsum,booktabs}
\usepackage{bm}

\newtheorem{theorem}{Theorem}

\theoremstyle{remark}

\theoremstyle{definition}

\newcommand{\argmin}{\mathop{\mathrm{argmin}}}

\def\V#1{\mathbf{#1}}
\def\C#1{\mathcal{#1}}

\def\B#1{\mathbb{#1}}
\def\S#1{\textsc{#1}}

\definecolor{Red}{rgb}{1,0,0}
\definecolor{Green}{rgb}{0,0.7,0}
\definecolor{Blue}{rgb}{0,0,1}
\definecolor{Red}{rgb}{0.6,0,0}
\definecolor{Orange}{rgb}{1,0.5,0}

\title{Nonlinear ISA with Auxiliary Variables for Learning Speech Representations}
\name{Amrith Setlur$^\dagger$, Barnabas Poczos$^\dagger$, Alan W Black$^\dagger$}
\address{
  $^\dagger$Carnegie Mellon University
  }
\email{asetlur@cs.cmu.edu, bapoczos@cs.cmu.edu, awb@cs.cmu.edu}

\begin{document}

\maketitle
\begin{abstract}
This paper extends recent work on nonlinear Independent Component Analysis \textsc{(ica)} by introducing a theoretical framework for nonlinear Independent Subspace Analysis \textsc{(isa)} in the presence of auxiliary variables. Observed high dimensional acoustic features like log Mel spectrograms can be considered as surface level manifestations of nonlinear transformations over individual multivariate sources of information like speaker characteristics, phonological content \textit{etc}. Under assumptions of energy based models we use the theory of nonlinear \S{isa} to propose an algorithm that learns  unsupervised speech representations whose subspaces are independent and potentially highly correlated with the original non-stationary multivariate sources. We show how nonlinear \S{ica} with auxiliary variables can be extended to a generic identifiable model for subspaces as well while also providing sufficient conditions for the identifiability of these high dimensional subspaces.  Our proposed methodology is generic and can be integrated with standard unsupervised approaches to learn speech representations with subspaces that can theoretically capture independent higher order speech signals. We evaluate the gains of our algorithm when integrated with the Autoregressive Predictive Decoding \S{(apc)} model by showing empirical results on the speaker verification and phoneme recognition tasks.
\end{abstract}

\input{sections/introduction}

\input{sections/theory}

\input{sections/proposed_methodology}
\input{sections/results}
\input{sections/conclusion}

\bibliographystyle{IEEEtran}
\bibliography{mybib}

\end{document}

%% file: sections/introduction.tex
\section{Introduction}
\label{sec:introduction}

The speech signals that we observe can be viewed as high-dimensional surface level manifestations of samples from independent non-stationary sources, that are entangled via a non-linear mixing mechanism. These sources can be entangled at session, utterance or segment levels \cite{hsu2017unsupervised}. Speech representations learnt by training deep recurrent models \cite{chung2019unsupervised, oord2018representation} over these surface level features fail to capture the original signals in their purest disentangled form. Unsupervised disentanglement of speech representations has been an active area of research \cite{li2018disentangled, chorowski2019unsupervised} since it has been shown that recovering independent factors of variation can improve the performance of downstream tasks like Automatic Speech Recognition (\S{asr}), especially under low resource constraints and domain mismatch \cite{hsu2017unsupervised}. Inspired by this, we propose an algorithm to learn unsupervised speech representations with independent subspaces, each of which can capture distinct disentangled source signals. These distinct subspaces can be potentially informative of patterns based on speaker characteristics or subphonetic events. This can be useful in learning a variety of acoustic models given very few labeled samples for each. 

Recently \cite{locatello2018challenging} it has been shown that learning disentangled representations is impossible without explicit bias on the algorithm and the data. Hence, we leverage a more principled approach to capturing the independent sources through the lens of nonlinear Independent Subspace Analysis (\S{isa}) in the presence of auxiliary variables. 


Nonlinear Independent Component Analysis (\S{ica}) is a provably unidentifiable problem \cite{hyvarinen1999nonlinear} as opposed to linear \S{ica} \cite{hyvarinen2000independent} which is identifiable given non-gaussian sources and other fundamental restrictions on the mixing matrix \cite{hyvarinen1999survey}. Attempts \cite{tan2001nonlinear, almeida2003misep, dinh2014nice} have been made to solve nonlinear \S{ica} for \textit{i.i.d} distributions under slightly stronger assumptions on the generative process \cite{ brakel2017learning, lee2004non}.   Recent progress in the field \cite{khemakhem2019variational,khemakhem2020ice} has revolved around a generic identifiable model that renders the latent sources conditionally independent in the presence of auxiliary variables. But most of the work \cite{hyvarinen2018nonlinear,hyvarinen2016unsupervised} has been focused on univariate sources which means that these models can't be directly applied to speech where the source signals are very high dimensional. Hence, we extend the auxiliary variables model proposed by \cite{hyvarinen2018nonlinear} for multivariate sources by \textit{first} stating sufficient conditions for the separability of sources and \textit{then}, providing training objectives suitable for learning speech representations with finite audio samples. Nonlinear \S{isa} is leveraged to learn unsupervised features on large unlabeled speech datasets. Using these features, simpler (linear) models are learnt on small labeled datasets.


Numerous approaches \cite{chorowski2019unsupervised,chung2018speech2vec, milde2018unspeech, hsu2017learning} have been proposed for learning unsupervised speech representations. Recent ones \cite{chung2019unsupervised,chorowski2019unsupervised} have been based on predictive coding schemes that use language model like objectives.
In parallel, there have been efforts to learn quantized representations via temporal segmentation and phonetic clustering \cite{liu2020towards} so as to map frame representations to linguistic units. But such models are fairly complicated and tricky to train. Also, most of these methods learn highly entangled representations that suffer from spurious correlations in the underlying data and thus fail to generalize. Our proposed algorithm improves upon these approaches by advocating for independent subspaces attained via additional constraints in the original optimization objectives. We begin by providing a theoretically identifiable model for nonlinear \S{isa} and then discuss how the model can be incorporated into existing methods for learning unsupervised speech representations. 




%% file: sections/theory.tex
\section{Theory}
\label{sec:theory}

We introduce a generative model of the observed data that we assume henceforth and present conditions under which, the original multi-dimensional sources are identifiable. We assume that the observed data $\V{x} \in \C{X} \subset \B{R}^{nd}$ is generated by applying a non-linear invertible transform $f$ on $n$ source signals $\V{s}_1 \dots \V{s}_n \in \C{S} \subset \B{R}^{d}$. We are given a dataset $\C{D} = \{(\V{x}^{(i)}, \V{u}^{(i)})\}_{i=1}^{N}$ with $N$ samples where each $\V{x}^{(i)} = f(\V{s}^{(i)})$ , $\V{s}^{(i)} = \bigotimes_{j=1}^{n} \V{s}_j^{(i)} = [\V{s}_1^{(i)} \dots \V{s}_n^{(i)}]$\footnote{Here $\bigotimes$ denotes the concatenation operation.}. Here, $\V{u}^{(i)} \in \C{U} \subset \B{R}^{p}$ denotes the corresponding auxiliary variable for $\V{x}^{(i)}$
and $f:\C{S}^{n} \rightarrow \C{X}$ is a non-linear mixing function (\textit{eqn.} \ref{eq:data-gen-1}), which is invertible and continuously differentiable almost everywhere (\textit{a.e}). The objective is to learn representations that can recover the source signals $(\{\V{s}_i\}_{i=1}^{n})$ up to an identifiability factor that we shall define shortly. For notational convenience, we denote the ${j}^{th}$ scalar element in a vector $\V{z}$ as $z_{j}$ and the $i^{th}$ consecutive $d$-dimensional vector ($i^{th}$ subspace) in  $\V{z}$ as $\V{z}_{i}$ or as $\V{z}_{i:} = \begin{bmatrix}z_{(i-1)d+1} \dots z_{id}\end{bmatrix}$.


\textbf{Model} The source distributions $\{p_i(\V{s}_i)\}_{i=1}^{n}$ are assumed to be independent given the auxiliary variable $\V{u}$ (\textit{eqn.} \ref{eq:data-gen-1}) and their densities are given by conditional energy based models (\textit{eqn.} \ref{eq:dist_den}) which have universal approximation capabilities \cite{khemakhem2020ice}. 
\begin{align}
    \label{eq:data-gen-1}
    & \V{x} = f(\V{s})  \quad\quad\quad  \log p(\V{s}|\V{u}) = \sum\limits_{i=1}^{n}  \log p_i(\V{s}_i|\V{u})  \\
    \label{eq:dist_den}
    & p_i(\V{s}_i|\V{u}) = \frac{\exp{\phi_i(\V{s}_i)^T\eta_i(\V{u})}}{Z_i(\V{u})} \quad \quad \begin{array}{c}\phi_i : \C{S} \rightarrow \B{R}^{m} \\ \eta_i : \C{U} \rightarrow \B{R}^m\\ \end{array}
\end{align}

\textbf{Definition of Identifiability} \label{def:id_def} We shall define the original sources $\{\V{s}_i\}_{i=1}^{n}$ to be identifiable if there exists an algorithm that takes as input a pair comprising of the observed sample and the corresponding auxiliary variable $(\V{x}=f(\V{s}), \V{u})$, and outputs $\begin{bmatrix}g_1(\V{s_{\pi_1}}), \dots, g_n(\V{s}_{\pi_n})\end{bmatrix}$, for some permutation $\pi:\B{N}^n \rightarrow \B{N}^n$ over $\{1 \dots n\}$. Each $g_i:\C{S} \rightarrow \C{S}$ is an invertible (\textit{a.e}) function and is defined as a function of a single distinct source $\V{s}_{\pi_i}$.

Popular algorithms \cite{hyvarinen2000independent} in linear \S{ica} \cite{hyvarinen1999survey} rely on estimators of Mutual Information (\S{mi}) to be able to separate the observed mixed samples into samples from the original source signals. Similarly, for nonlinear \S{ica} we compute \S{mi} between the observed and auxiliary variables  ($\V{I}(\V{x}, \V{u})$ in \textit{eqn.} \ref{eq:nce-1}) using Noise Contrastive Estimation \textsc{(nce)} \cite{gutmann2012noise}. A nonlinear logistic classifier is used to distinguish between correct (observed) pairs $(\V{x}^{(i)}, \V{u}^{(i)})$ and randomly generated incorrect pairs $(\V{x}^{(i)}, \tilde{\V{u}}^{(i)})$ where $\tilde{\V{u}}^{(i)}$ is drawn from the marginal distribution over $\V{u}$. The regression function for this logistic classifier is given by $r(\V{x}, \V{u})$, where $h_i: \C{X} \rightarrow \B{R}^{d}\,, \psi_i:\B{R}^{d} \times \B{R}^{p} \rightarrow \B{R} \in L^{2}$ are sufficiently smooth universal function approximators (neural networks) and $\forall i\,, h_i$ is invertible \textit{a.e}.
 \begin{align}
    \V{I}(\V{x}, \V{u}) = \int_{x,u}  \log  \frac{p(\V{x}, \V{u})}{p(\V{x}) p(\V{u})}  &\, d\B{P}(\V{x}, \V{u}) = \int  r(\V{x}, \V{u}) \, d\B{P}(\V{x}, \V{u}) \nonumber \\
    \textrm{\textit{where,}} \quad r(\V{x}, \V{u}) &= \sum_{i=1}^n \psi_i(h_i(\V{x}), \V{u})  \label{eq:nce-1}
\end{align}
The following main \S{isa} separation theorem states that the vector $h(\V{x})=\bigotimes_{i=1}^{n} h_i(\V{x}) \in \B{R}^{nd}$, with subspaces $h_i(\V{x}) \in \B{R}^{d}$ can recover  $\V{s}_i$ since $\exists \pi, \{g_i\}_{i=1}^{n}$ such that $h_i(\V{x})=g_i(\V{s}_{\pi_i})$.

\begin{theorem}
Given that we observe the dataset $\C{D}$ with $N$ samples: $\{\V{x}^{(i)} = f(\V{s}^{(i)}), \V{u}^{(i)}\}_{i=1}^{N}$ generated by a model based on \textit{eqns.} (\ref{eq:data-gen-1}, \ref{eq:dist_den}), then under the following assumptions\footnote{  The \textbf{separability} \textit{assm.} requires the auxiliary variables $\V{u}$ to have a sufficiently strong and diverse effect on the source distributions \cite{hyvarinen2018nonlinear}.}:
\begin{enumerate}
    \item \textbf{Realizability Assumption:}  Given infinite ($N \rightarrow \infty$) samples one can efficiently learn $\psi_i^*, h_i^*$ such that the \textsc{nce} algorithm can estimate the mutual information $\V{I}(\V{x}, \V{u})$ with an arbitrarily small error, using the regression function $r(\V{x}, \V{u})$ which follows the form in \textit{eqn.} \ref{eq:nce-1}. 
    \item \textbf{Separability Assumption:} $\forall \V{s} \in \C{S}^n$,  $\V{z} \neq 0 \in \B{R}^{d}$ with first and second order derivatives given by tensors $\nabla  \phi_i(\V{s}_i)$ $\in$ $\B{R}^{m\times d}$ and $\nabla^2$  $\phi_i(\V{s}_i)$ $\in$ $\B{R}^{m\times d\times d}$ respectively; $\exists \{\V{u}_l\}_{l=0}^{2nd}$ $\in$ $\C{U}^{2nd+1}$ such that: $$ \left\{ \bigotimes_{i=1}^{n} \left( \begin{bmatrix}\nabla  \phi_i(\V{s}_i)^T \\ (\nabla^2  \phi_i(\V{s}_i) \, \bar{\times}_3 \footnotemark \, \V{z})^T\end{bmatrix} \zeta_i(\V{u}_l, \V{u}_0) \right)\right\}_{l=1}^{2nd}$$ spans $\B{R}^{2nd}$ for $\zeta_i(\V{u}_l, \V{u}_0)=(\eta_i(\V{u}_l) - \eta_i(\V{u}_0))$,
    \footnotetext{$\bar{\times}_3$ denotes the $3^{rd}$ mode product \cite{kolda2009tensor}.}
\end{enumerate}
the subspaces $\{h_i(\V{x})\}_{i=1}^{n}$ can identify the conditionally independent sources $\{\V{s}_i\}_{i=1}^{n}$ up to the \textbf{definition of identifiability}.  
\end{theorem}
\textit{Proof Sketch\footnote{For more details on the validity and necessity of similar results for independent components (as opposed to subspaces) we refer the reader to \cite{hyvarinen2018nonlinear}. Also, for the sake of completion we show a proof sketch for the identifiability of our \S{isa} model. It's an extension of the proof for the univariate case
\cite{hyvarinen2018nonlinear,hyvarinen2016unsupervised}.}:} For an observed sample $\V{x} \in \C{X}$, let $\V{y}=h^*(\V{x})$ be given by the optimal functions $\{h_i^*\}_{i=1}^n$. The functions $\{\psi_i^*, h_i^*\}_{i=1}^{n}$\footnote{Subscript $i$ is dropped wherever it can be understood from context.} are learnt using the \S{nce} objective whose regression function is given by $r(\V{x}, \V{u})$. Since $\V{s} = f^{-1}(h^{-1}(\V{y}))$ is a composition of two invertible transforms, we introduce  $v:\B{R}^{nd} \rightarrow \C{S}^n$ where $\V{s}=v(\V{y})$. Also, let $f^{-1}$ be denoted by $g$. From \textit{eqn.} \ref{eq:nce-1} we know that $r(\V{x}, \V{u}) = \log p(\V{x}| \V{u}) - \log p(\V{x})$,

Using the density transformation rules \cite{murphy2012machine} for invertible functions we can show that, $\log p(\V{x}| \V{u})$ $=$ $\log p(\V{s}|\V{u})$ $+$ $\log |\det \V{J}g(\V{x})|$ and $\log p(\V{x})$ $=$ $\log p(\V{s})$ $+$ $\log |\det \V{J}g(\V{x})|$ . Thus, $r(\V{x}, \V{u}) = \log p(\V{s}|\V{u}) - \log p(\V{s})$. Using \textit{eqn.} \ref{eq:nce-1}:
\begin{align}
\label{eqn:pf-11}
\sum_{i=1}^n \psi_i^*(\V{y}_i, \V{u}) &=  \log p(v(\V{y})|\V{u}) - \log p(v(\V{y}))
\end{align}
We begin by substituting \textit{eqns.} \ref{eq:data-gen-1}, \ref{eq:dist_den} in the above result. Also, since \textit{eqn.} \ref{eqn:pf-11} holds true for $ \{\V{u}_l\}_{l=0}^{2nd}$, we can get $2nd+1$ such equations and from each we can subtract the equation given by $\V{u}_0$, which leaves us with $2nd$ \textit{eqns.} of the form $\sum_{i=1}^{n} \phi_i(v(\V{y})_{i:})^T\zeta_i(\V{u}_l, \V{u}_0) - (\log Z_i(\V{u}_l) - \log Z_i(\V{u}_0)) =  \sum_{i=1}^n \psi_i^*(\V{y}_i, \V{u})$.   
Taking the derivative of both sides of this \textit{eqn.} \textit{w.r.t.} $y_j$ and subsequently w.r.t $y_k$ \textit{s.t.} $\lceil j/d \rceil \neq \lceil k/d \rceil$ we get,
\begin{align}
    \label{eqn:pf-12}
    & 0 =  \sum_{i=1}^{n} \left(\underbrace{\nabla  \phi_i(v(\V{y})_{i:})}_{\text{\textcircled{1}}} \frac{\partial^2 v(\V{y})_{i:}}{\partial y_j \partial y_k}\right)^T   \zeta_i(\V{u}_l, \V{u}_0) \nonumber \\
    & +  \left(\underbrace{\left(\nabla^2 \phi_i(v(\V{y})_{i:}) \bar{\times}_3 \frac{\partial v(\V{y})_{i:}}{\partial y_j}\right)}_{\text{\textcircled{2}}} \frac{\partial v(\V{y})_{i:}} {\partial y_k}\right)^T \zeta_i(\V{u}_l, \V{u}_0) \nonumber
\end{align}
Concatenating \textcircled{1}, \textcircled{2} into a single matrix in $\B{R}^{2d \times m}$, the above can be written as a single euclidean inner product in $\B{R}^{2nd}$.
\begin{align}
    & \left(\bigotimes_{i=1}^{n} \left( \begin{bmatrix}\nabla  \phi_i(\V{s}_i)^T \\ \left(\nabla^2  \phi_i(\V{s}_i) \, \bar{\times}_3 \, \frac{\partial v(\V{y})_{i:}}{\partial y_j}\right)^T\end{bmatrix}\zeta_i(\V{u}_l, \V{u}_0) \right)\right) \Gamma(\V{y}) = 0 \nonumber
\end{align}
For $\Gamma(\V{y})=\left(\bigotimes\limits_{i=1}^{n}  \left[ \frac{\partial^2 v(\V{y})_{i:}}{\partial y_j \partial y_k}  \frac{\partial v(\V{y})_{i:}}{\partial y_k} \right] \right) \in \B{R}^{2nd}$  the above equation holds true for $2nd$ distinct values of the auxiliary variable $\V{u}_l$. For invertible $v$, if we assume that $\frac{\partial v(\V{y})_{i:}}{\partial y_j} \neq 0$ then we can apply the \textbf{separability} \textit{assm.} which implies $\Gamma(\V{y}) = \V{0}$. This further implies that $\frac{\partial v(\V{y})_{i:}}{\partial y_k} = 0$. Thus $\forall i$, $\frac{\partial v(\V{y})_{i:}}{\partial y_j} \lor \frac{\partial v(\V{y})_{i:}}{\partial y_k}$. Since $\lceil j/d \rceil \neq \lceil k/d \rceil$, $y_j$ and $y_k$ belong to distinct subspaces of $y=h(\V{x})$. Hence the $i^{th}$ source given by $v(\V{y})_{i:}$ cannot simultaneously be a function of two distinct subspaces of $h(\V{x})$. Given the invertible function $f(h(\cdot))$ with its full rank jacobian we can recover the sources $\{\V{s}_i\}_{i=1}^n$ via the subspaces of $\V{h(x)}$;  $h_i(\V{x})=g_i(\V{s}_{\pi_i})$ for an invertible function $g_i$, permutation $\pi$. 





\textbf{Hilbert-Schmidt Independence Criterion (\textsc{hsic})}  \cite{gretton2008kernel} The above theorem proves the existence of functions $\psi^*, h^*$ that can not only compute  $\V{I}(\V{x}, \V{u})$ with arbitrary precision but can also recover the original multi-dimensional sources. Albeit, \S{nce} algorithm relies on the assumption of infinite samples of positive $(\V{x}, \V{u})$ and negative $(\V{x}, \V{\tilde{u}})$ pairs which is rarely true in practice. Hence, along with the \S{nce} objective which learns $r(\V{x}, \V{u})$ that distinguishes between those pairs, we introduce constraints imposed via the \S{hsic} estimator that specifically accounts for independence amongst the subspaces of $h(\V{x})$. This acts as a strong inductive bias to learn $\psi^*, h^*$ with finite observed samples of $(\V{x}, \V{u})$. \S{hsic} is a kernel based statistical test of independence for two multivariate random variables and is well suited for high dimensional data as opposed to tests \cite{read2012goodness, kankainen1995consistent, feuerverger1993consistent}  based on the power divergence family and characteristic functions which are mainly meant for low-dimensional random variables \cite{gretton2008kernel}. 
Given $\C{D} = \{\V{x}^{(i)}, \V{u}^{(i)}\}_{i=1}^{N}$ with $N$ samples, let the set of features $(h(\V{x}^{(i)}))$ be denoted by $\{\V{y}^{(i)}=h(\V{x}^{(i)})\}_{i=1}^{N}$.
For $\B{R}^d$ dimensional subspaces $j,k$ let $\V{y}_j \in \C{Y}_j \subseteq \B{R}^d,\,\V{y}_k \in \C{Y}_k \subseteq \B{R}^d$ and $\V{P}_{jk}$ denote a Borel probability measure over $\C{Y}_j \times \C{Y}_k$ with $N$ i.i.d samples $\C{Z}_{jk}:=\{(\V{y}_{j}^{(i)}, \V{y}_{k}^{(i)})\}_{i=1}^{N}$ drawn from it.  If $\C{F}, \C{G}$ are two Reproducible Kernel Hilbert Spaces \textsc{(rkhs)} equipped with kernels\footnotemark $\,\,k_f, k_g$  then the biased empirical \S{hsic} criterion $\B{\hat{H}}_{jk} = \frac{1}{N^2} \textrm{tr}(\V{K_f}^{(j)}\V{H}\V{K_g}^{(k)}\V{H})$ and $\V{K_f}^{(j)}[p,q]=k_f(\V{y}_{j}^{(p)}, \V{y}_{j}^{(q)})$, $\V{K_g}^{(k)}[p,q]=k_g(\V{y}_{k}^{(p)}, \V{y}_{k}^{(q)})$,   $\V{H} = \V{I}-\frac{1}{N}\V{1}\V{1}^T \in \B{R}^{N \times N}$.  
\footnotetext{$k_f:\C{Y}_j \times \C{Y}_j \rightarrow \B{R}$, $k_g:\C{Y}_k \times \C{Y}_k \rightarrow \B{R} $; for $\V{z},\V{z'}\in \C{Y}_j$, $k_f(\V{z}, \V{z'})=\langle k_f(\V{z}, \cdot), k_f(\V{z'}, \cdot) \rangle_{\C{F}}$ and for  $\V{z},\V{z'} \in \C{Y}_k$, $k_g(\V{z}, \V{z'})=\langle k_g(\V{z}, \cdot), k_g(\V{z'}, \cdot) \rangle_{\C{G}}$.} 
    

\textbf{Algorithm (\textsc{nce-hsic})} We have shown that the \S{nce} algorithm can learn a regression function of the form $r(\V{x}, \V{u})$ (\textit{eqn.} \ref{eq:nce-1}) with optimal predictors $\psi^*, h^*$ such that the subspaces of $h^*(\V{x})$ can recover the original sources $\V{s}_i$.  Constrained by a finite dataset we use the biased empirical \S{hsic} estimator $\B{\hat{H}}_{jk}$ (lower values imply more independence) as an additional objective while optimizing for $\psi^*, h^*$. If the true and noisy samples for the \textsc{nce} algorithm are given by $(\V{x}^{(l)}, \V{u}^{(l)})$ and $(\V{x}^{(l)}, \V{u}^{(l'\neq l)})$ respectively, then the final loss objective $\C{L}_{nh}$ for \S{nce-hsic} is:
\begin{align}
    \C{L}_{nh} = \frac{1}{N} \sum_{l \in [N]}  r(\V{x}^{(l)}, \V{u}^{(l'\neq l)}) - r(\V{x}^{(l)}, \V{u}^{(l)}) + \lambda \sum_{j,k} \B{\hat{H}}_{jk} \nonumber
\end{align}

%% file: sections/proposed_methodology.tex
\section{Proposed Methodology}
\label{sec:proposed_methodology}



Speech representations that can explicitly capture factors of variation like phoneme identities or speaker traits while being invariant to other factors like underlying pitch contour or background noise \cite{li2018disentangled, chorowski2019unsupervised} have  proven to be beneficial since they are less prone to overfitting on spurious correlations in the data. Nevertheless, disentanglement is hard to achieve in general due to the presence of confounding variables \cite{locatello2018challenging}. In this section, we introduce our approach \textbf{\textsc{apc-nce-hsic}} or \textbf{\S{anh}} to learn representations with independent subspaces that can theoretically capture distinct acoustic/linguistic units relevant for downstream tasks like \S{asr}. 

Nonlinear \S{isa} provides us with a simple yet principled framework for learning speech representations in the presence of auxiliary variables, which in the case of sequential data like speech can be ``time". Learning unsupervised representations can be posed as a problem of recovering from entangled samples the non-stationary sources that are independent given the auxiliary variable (time frame sequence). The \S{nce-hsic} algorithm can be used to identify original factors of variation via distinct independent subspaces.  In order to ensure that the independent subspaces are not only mutually exclusive but are also having a high \S{mi} with surface features like Mel-frequency cepstral coefficients (\S{mfcc}) or log Mel spectrograms (\S{lms}) we build on existing approaches based on predictive coding strategies \cite{chung2018speech2vec, oord2018representation}. Although our algorithm can be seamlessly integrated  into any of these methods, in this work we show empirical results that highlight the performance improvements gained by incorporating the \S{nce-hsic} criterion into the \S{apc} model.

\textbf{\S{apc}} \cite{chung2019unsupervised} is a language model based method to learn unsupervised speech representations. It uses a Recurrent Neural Network (\textsc{rnn}) to model temporal information within an acoustic sequence comprising of $80$-dimensional \textsc{lms} features $\{\V{x}_i\}_{i=0}^{T}$. Given these features until a fixed time step $t$, the \textsc{apc} model predicts the surface feature $\tau$ time steps ahead \textit{i.e.} $\V{x}_{t+\tau}$. If $\{\V{\hat{p}}_i\}_{i=0}^{T-\tau}$ represents the sequence predicted by the \textsc{rnn}, then the $l_1$ loss used to train the model is given by: 
\begin{align}
    \C{L}_{apc}(\V{x}) = \sum_{i=0}^{T-\tau} -\log p(\V{x}_{i+\tau}|\V{x}_1\dots \V{x}_i) = \sum_{i=0}^{T-\tau} |\V{\hat{p}}_i - \V{x}_{i+\tau}| \nonumber
\end{align}
\textbf{\textsc{apc-nce-hsic} or \S{anh}} is our proposed model where features with independent subspaces are learnt through the $\S{nce-hsic}$ criterion which is applied to the hidden states of the \S{rnn} module trained with the \S{apc} objective above. Specifically, the function $h(\V{x})$ is modeled using the \S{rnn}. The \S{nce-hsic} criterion increases the correlation of the original sources with the subspaces of $h(\V{x})$ or in this case the subspaces of the hidden states of the \S{rnn}.  If the \S{rnn} is parameterized by $\theta \in \Theta$ then the hidden state can be represented as the function $h(\theta, \V{x})$. With $r(\V{x}, \V{u}) = \sum_{i=1}^n \psi_i(h_i(\theta, \V{x}), \V{u})$ the final objective would be:
\begin{align}
  \argmin_{\{\psi_i\}_{i=1}^n, \theta}  \C{L}_{anh} = \frac{1}{|\C{D}|}\sum_{\V{x} \in \C{D}}\C{L}_{apc} (\V{x}) + \beta \C{L}_{nh}
\end{align}
\textbf{Auxiliary Variables}\footnote{Auxiliary variables can be potentially given by other domains like the frequency spectrum, but in this work we focus only on time.} The original \textsc{lms} sequence of length $T$ is fragmented into time segments $\{s_j\}_{j=1}^{\lceil T/\gamma \rceil}$ of length $\gamma$, and each element $\V{x}_{j,t}$ in a given segment $s_j$ has its auxiliary variable $\V{u}_{j,t}$ set to the value $j$, which is nothing but the corresponding segment's position in the input sequence. The hidden states of the \S{rnn} along with the generated auxiliary variables are passed to the \textsc{nce} module which \textit{first}, generates positive $(\V{x}_t, \V{u}_t)$ and negative $(\V{x}_t, \V{\tilde{u}}_t)$ pairs and \textit{then}, learns $\psi^*, \theta^*$ to distinguish between them optimally. Upon the commencement of the unsupervised learning phase, the hidden state for the $t^{th}$ frame with surface features $\V{x}_t$ would comprise of $n$ subspaces ($\{h_i(\theta^*, \V{x}_t)\}_{i=1}^n$) that capture different factors of variation, independent for the same value of the auxiliary variable $\V{u}_t$. Thus the hidden states can efficiently decouple factors that vary independently locally.




\S{nce} is a powerful tool to predict \S{mi} and has been used in recent works like \textbf{\S{cpc}} \cite{oord2018representation} that rely on the \S{nce} objective to distinguish pairs of context vectors from the same or different time segments. This approach is similar to Time Contrastive Learning \S{tcl} \cite{hyvarinen1999nonlinear} which is an algorithm for nonlinear \S{ica}. Although \S{tcl} has only been shown to work for univariate cases and \S{cpc} fails to model independent subspaces explicitly, they serve as a strong motivation for our approach which addresses both concerns. 





%% file: sections/results.tex
\section{Experiments and Results}
\label{sec:experiments-and-results}

In this section, we empirically evaluate the performance of the proposed \textsc{anh} algorithm against two baseline models: \S{apc} and \S{cpc} on two downstream tasks, (1) phoneme recognition (\S{pr}) and (2) speaker verification (\S{sv}).

\textbf{Datasets and Implementation} LibriSpeech corpus \cite{panayotov2015librispeech} was used for unsupervised training of the \S{anh} model and other baselines. The datasets for \S{pr} and \S{sv} were picked from \S{wsj} \cite{paul1992design} and \S{timit} corpora respectively \footnote{For brevity we skip the details of the dataset and refer the reader to \cite{chung2019unsupervised} from where we borrowed the dataset splits and input \S{lms} features.}. For \S{apc} we use a multi-layer unidirectional \S{lstm} with residual connections exactly as detailed in \cite{chung2019unsupervised}, with the exception of using 4 layers in the \S{lstm} (wherever mentioned explicitly) and for \S{cpc} modifications suggested in \cite{chung2019unsupervised} are made for a fair comparison. In the \textit{unsupervised} phase we train the \S{rnn} using the $\C{L}_{anh}$ objective. The \S{rnn} hidden states which are $512$-dimensional are assumed to be a collection of $n=4$ contiguous subspaces each of which has $d=128$ dimensions. These $4$ subspaces of the \S{rnn} parameterized by $\theta$, represent the output $\{h_i(\theta, \V{x})\}_{i=1}^{4}$ where $\V{x}$ is the \S{lms} feature and $h_i(\theta, \V{x})$ is the $i^{th}$ subspace. The \S{nce} module also needs $\psi_i(\cdot, \cdot)$ which is implemented using 4-layer \textsc{mlp}s, with ReLU activations, dropouts and batch-normalization. For $\C{L}_{nh}$\footnote{The optimal $\beta$ in $\C{L}_{anh}$ \& $\lambda$ in $\C{L}_{nh}$ were found to be $0.1$ and $0.02$.}, five negative pairs are drawn for every positive pair. In the \textit{supervised} phase, once the \textsc{isa} features ($h(\theta^*, \V{x})$) given by the hidden states (final layer) of the trained \textsc{rnn} ($\theta^*$)  are extracted, a supervised linear classifier is trained over features from each frame for \S{pr} whereas an \S{lda} model is trained over features averaged over the entire sequence for \S{sv}.



\textbf{Phoneme Recognition} Table \ref{tab:libri-phone-per} highlights the performance (Phone Error Rates \S{(per)}) of our approach (\S{anh})\footnote{Unless specified all \S{anh} models are trained with $\gamma=30$.} against the best variants of the \S{cpc}, \S{apc} models. The supervised baseline (\S{lms}+\S{mlp}) which involves training a 3-layer nonlinear classifier over the \S{lms} features fails to capture contextual information. Even though \S{cpc} can learn contextual features, it only captures information relevant for recognizing contexts that are $\tau$ steps apart. Thus it may ignore signals that remain relatively stationary for the entire utterance \cite{chung2019unsupervised}. On the other hand \S{apc} directly predicts surface features $\tau$ steps ahead and thus can model sub-phonetic context useful in predicting the next phone. \S{anh} with $\tau=5$ has the least \S{per} since the addition of the \S{nce-hsic} objective enables the model to learn noise-free subspaces that can capture relevant factors like formant movements. Finally, adding layers to the \S{rnn} further improves the scores. 
\begin{table}[!htbp]
  \caption{Performance comparison (based on PER) on the Phoneme Recognition task (\S{wsj} corpus \cite{paul1992design}).
}
  \label{tab:libri-phone-per}
  \centering
  \begin{tabular}{c c c c}
    \toprule
    \textbf{Method} &  \multicolumn{3}{c}{\textbf{PER}} \\ \toprule
    \# lookahead-steps ($\tau$) & 2 & 5 & 10 \\\midrule
    \S{lms}+\textsc{mlp} (supervised)     &    \multicolumn{3}{c}{42.5}     \\
      \textsc{cpc} \cite{oord2018representation}                     &    41.8  & 44.6&  47.3  \\
      \textsc{apc} (3-layer) \cite{chung2019unsupervised}                &    36.6  & 35.7 & 35.5  \\
      \textsc{apc} (4-layer) \cite{chung2019unsupervised} & 34.5  & 35.2 & \textbf{33.8} \\
      \textsc{anh} (3-layer) (Ours) &  33.2 & \textbf{31.3} & 34.7    \\
      \textsc{anh} (4-layer) (Ours) &   \textbf{31.9} & 31.8  & 34.2    \\
    \midrule
    \textbf{Ablations}  & 2 & 5 & 10  \\\midrule
    \textsc{apc} + \S{nce} &  32.0 & 32.4  & 34.3    \\
    \S{nce-hsic} &   49.8 & 48.5  &  54.6 \\
    \S{nce} & 49.4 & 53.2 & 55.9 \\\bottomrule
  \end{tabular}
 \end{table}

\textbf{Speaker Verification} Results for \S{sv} are summarized in table \ref{tab:libri-speaker-acc} which shows lower Equal Error Rates (\S{eer}) achieved by \S{anh} as compared to the baselines. It has been shown that in deep language models, lower layers model local syntax while the higher ones capture semantic content \cite{chung2019unsupervised,peters2018dissecting}. We make similar observations since the \S{eer} values increase (for all $\tau$) when the \S{anh} model has more than 3 layers. Lowering $\tau$ reduced \S{eer} in most cases and had minimal impact on the independence ($\B{\hat{H}}_{jk}$).
 \begin{table}[!htbp]
  \caption{Performance (based on EER) on the speaker verification task (\S{timit} corpus). ($^*$choosing different layers \cite{chung2019unsupervised})
  }
  \label{tab:libri-speaker-acc}
  \centering
  \begin{tabular}{c c c c c}
    \toprule
    \textbf{Method} &  \multicolumn{4}{c}{\textbf{\S{EER}}} \\\toprule
    \# lookahead-steps ($\tau$) & 2 & 3 & 5 & 10 \\\midrule
      \textsc{cpc} features \cite{oord2018representation} & 5.62 &  5.29      &   5.42   &  6.01 \\
      \textsc{apc} (3-layer)-1$^*$ \cite{chung2019unsupervised} & 3.82 & 3.67      &   3.88     & 4.01  \\
      \textsc{apc} (3-layer)-2$^*$  \cite{chung2019unsupervised}&   \textbf{3.41}   & 3.72   &   3.92     & 4.04 \\
       \textsc{anh} (2-layer) (Ours) &  3.53 &    3.35 & 3.91 & 4.12 \\
       \textsc{anh} (3-layer) (Ours) & 3.45    & \textbf{3.12}     & \textbf{3.45} & \textbf{3.67}\\ 
    \bottomrule
  \end{tabular}
  \end{table}

\textbf{Ablations} \S{nce-hsic} model when trained without the $\C{L}_{apc}$ loss rendered independent subspaces but performed poorly on \S{pr} since there is no reason to believe why such subspaces would retain phonetic information. Adding the \S{apc} objective aids the model (\S{anh}) to learn acoustic features while disentangling the factors across subspaces (table \ref{tab:libri-phone-per}). Removing the \S{hsic} criterion increased the \S{per} and the model training also took ($\times$2) longer to converge. This reinforces our hypothesis that the \S{hsic} criterion provides a good inductive bias for a more generalizable model.

\textbf{Independence} In order to measure the independence of the four $128$-dimensional subspaces of the \S{rnn} states, absolute values of the Pearson's Correlation were computed on the validation splits for \S{pr,sv}. When averaged over all possible pairs, they were found to be $0.21$, $0.19$ on \S{pr,sv} respectively when both \S{nce} and \S{hsic} objectives were considered in $\C{L}_{nh}$. With $\lambda = 0$ these values were $0.29$ and $0.33$ but were still significantly lower as compared to the case of \S{apc} which had average absolute correlation values of $0.81$ and $0.77$ on \S{pr} and \S{sv} respectively.

\textbf{Time Segment Length ($\gamma$)} We show the impact of the time segment length $\gamma$ on the phoneme classification task in table \ref{tab:libri-other}. As we increase $\gamma$ the total number of segments (and auxiliary variables) reduce in an utterance. Theoretically, $2nd$ distinct auxiliary variables are needed to identify $n$ sources each of which is $d$-dimensional (\textit{sec.} \ref{sec:theory}). Hence increasing $\gamma$ to values greater than $50$ leads to higher $(>40)$ \S{per}s. Additionally, we observe that when the \S{rnn} is trained with higher values of $\tau$ for the \S{apc} objective \S{per} drops by using wider segments. This may indicate that the distribution of the underlying factors remain stationary for longer periods at higher values of $\tau$.  
 \begin{table}[!htbp]
  \caption{Comparing different values of ($\gamma$) for \S{anh} (3-layer) model on the phoneme classification task.
}
  \label{tab:libri-other}
  \centering
  \begin{tabular}{c c c c }
    \toprule
    \textbf{Segment size $\gamma$} &  \multicolumn{3}{c}{\textbf{PER}} \\ \toprule
    \# lookahead-steps ($\tau$) & 2 & 5 & 10 \\\midrule
      $\gamma=10$ &   39.4  & 38.5&  36.8  \\
      $\gamma=20$ &   38.1  & 35.3&  37.5  \\
      $\gamma=30$ &   \textbf{33.2}  & \textbf{31.3} &  34.7  \\
      $\gamma=50$ &   34.0  & 32.0&  \textbf{33.5} \\
      \bottomrule
  \end{tabular}
 \end{table}

%% file: sections/conclusion.tex
\section{Conclusion}
\label{sec:conclusion}
We extend nonlinear \S{ica} and show how the proposed algorithm to compute \S{mi} between the observed and auxiliary variables can provably identify independent subspaces under certain regularity conditions. We also use the algorithm to learn unsupervised speech representations with disentangled subspaces when integrated with existing approaches like \S{apc}. Future work may involve a close analysis of the features in these subspaces to understand which orthogonal components are represented by each and how they can prove to be useful for downstream tasks.

%% file: main.bbl
\begin{thebibliography}{10}
\providecommand{\url}[1]{#1}
\csname url@samestyle\endcsname
\providecommand{\newblock}{\relax}
\providecommand{\bibinfo}[2]{#2}
\providecommand{\BIBentrySTDinterwordspacing}{\spaceskip=0pt\relax}
\providecommand{\BIBentryALTinterwordstretchfactor}{4}
\providecommand{\BIBentryALTinterwordspacing}{\spaceskip=\fontdimen2\font plus
\BIBentryALTinterwordstretchfactor\fontdimen3\font minus
  \fontdimen4\font\relax}
\providecommand{\BIBforeignlanguage}[2]{{%
\expandafter\ifx\csname l@#1\endcsname\relax
\typeout{** WARNING: IEEEtran.bst: No hyphenation pattern has been}%
\typeout{** loaded for the language `#1'. Using the pattern for}%
\typeout{** the default language instead.}%
\else
\language=\csname l@#1\endcsname
\fi
#2}}
\providecommand{\BIBdecl}{\relax}
\BIBdecl

\bibitem{hsu2017unsupervised}
W.-N. Hsu, Y.~Zhang, and J.~Glass, ``Unsupervised learning of disentangled and
  interpretable representations from sequential data,'' in \emph{Advances in
  neural information processing systems}, 2017, pp. 1878--1889.

\bibitem{chung2019unsupervised}
Y.-A. Chung, W.-N. Hsu, H.~Tang, and J.~Glass, ``An unsupervised autoregressive
  model for speech representation learning,'' \emph{arXiv preprint
  arXiv:1904.03240}, 2019.

\bibitem{oord2018representation}
A.~v.~d. Oord, Y.~Li, and O.~Vinyals, ``Representation learning with
  contrastive predictive coding,'' \emph{arXiv preprint arXiv:1807.03748},
  2018.

\bibitem{li2018disentangled}
Y.~Li and S.~Mandt, ``Disentangled sequential autoencoder,'' \emph{arXiv
  preprint arXiv:1803.02991}, 2018.

\bibitem{chorowski2019unsupervised}
J.~Chorowski, R.~J. Weiss, S.~Bengio, and A.~van~den Oord, ``Unsupervised
  speech representation learning using wavenet autoencoders,'' \emph{IEEE/ACM
  transactions on audio, speech, and language processing}, vol.~27, no.~12, pp.
  2041--2053, 2019.

\bibitem{locatello2018challenging}
F.~Locatello, S.~Bauer, M.~Lucic, G.~R{\"a}tsch, S.~Gelly, B.~Sch{\"o}lkopf,
  and O.~Bachem, ``Challenging common assumptions in the unsupervised learning
  of disentangled representations,'' \emph{arXiv preprint arXiv:1811.12359},
  2018.

\bibitem{hyvarinen1999nonlinear}
A.~Hyv{\"a}rinen and P.~Pajunen, ``Nonlinear independent component analysis:
  Existence and uniqueness results,'' \emph{Neural Networks}, vol.~12, no.~3,
  pp. 429--439, 1999.

\bibitem{hyvarinen2000independent}
A.~Hyv{\"a}rinen and E.~Oja, ``Independent component analysis: algorithms and
  applications,'' \emph{Neural networks}, vol.~13, no. 4-5, pp. 411--430, 2000.

\bibitem{hyvarinen1999survey}
A.~Hyvarinen, ``Survey on independent component analysis,'' \emph{Neural
  computing surveys}, vol.~2, no.~4, pp. 94--128, 1999.

\bibitem{tan2001nonlinear}
Y.~Tan, J.~Wang, and J.~M. Zurada, ``Nonlinear blind source separation using a
  radial basis function network,'' \emph{IEEE transactions on neural networks},
  vol.~12, no.~1, pp. 124--134, 2001.

\bibitem{almeida2003misep}
L.~B. Almeida, ``Misep--linear and nonlinear ica based on mutual information,''
  \emph{Journal of Machine Learning Research}, vol.~4, no. Dec, pp. 1297--1318,
  2003.

\bibitem{dinh2014nice}
L.~Dinh, D.~Krueger, and Y.~Bengio, ``Nice: Non-linear independent components
  estimation,'' \emph{arXiv preprint arXiv:1410.8516}, 2014.

\bibitem{brakel2017learning}
P.~Brakel and Y.~Bengio, ``Learning independent features with adversarial nets
  for non-linear ica,'' \emph{arXiv preprint arXiv:1710.05050}, 2017.

\bibitem{lee2004non}
J.~A. Lee, C.~Jutten, and M.~Verleysen, ``Non-linear ica by using isometric
  dimensionality reduction,'' in \emph{International Conference on Independent
  Component Analysis and Signal Separation}.\hskip 1em plus 0.5em minus
  0.4em\relax Springer, 2004, pp. 710--717.

\bibitem{khemakhem2019variational}
I.~Khemakhem, D.~P. Kingma, and A.~Hyv{\"a}rinen, ``Variational autoencoders
  and nonlinear ica: A unifying framework,'' \emph{arXiv preprint
  arXiv:1907.04809}, 2019.

\bibitem{khemakhem2020ice}
I.~Khemakhem, R.~P. Monti, D.~P. Kingma, and A.~Hyv{\"a}rinen, ``Ice-beem:
  Identifiable conditional energy-based deep models,'' \emph{arXiv preprint
  arXiv:2002.11537}, 2020.

\bibitem{hyvarinen2018nonlinear}
A.~Hyvarinen, H.~Sasaki, and R.~E. Turner, ``Nonlinear ica using auxiliary
  variables and generalized contrastive learning,'' \emph{arXiv preprint
  arXiv:1805.08651}, 2018.

\bibitem{hyvarinen2016unsupervised}
A.~Hyvarinen and H.~Morioka, ``Unsupervised feature extraction by
  time-contrastive learning and nonlinear ica,'' in \emph{Advances in Neural
  Information Processing Systems}, 2016, pp. 3765--3773.

\bibitem{chung2018speech2vec}
Y.-A. Chung and J.~Glass, ``Speech2vec: A sequence-to-sequence framework for
  learning word embeddings from speech,'' \emph{arXiv preprint
  arXiv:1803.08976}, 2018.

\bibitem{milde2018unspeech}
B.~Milde and C.~Biemann, ``Unspeech: Unsupervised speech context embeddings,''
  \emph{arXiv preprint arXiv:1804.06775}, 2018.

\bibitem{hsu2017learning}
W.-N. Hsu, Y.~Zhang, and J.~Glass, ``Learning latent representations for speech
  generation and transformation,'' \emph{arXiv preprint arXiv:1704.04222},
  2017.

\bibitem{liu2020towards}
A.~H. Liu, T.~Tu, H.-y. Lee, and L.-s. Lee, ``Towards unsupervised speech
  recognition and synthesis with quantized speech representation learning,'' in
  \emph{ICASSP 2020-2020 IEEE International Conference on Acoustics, Speech and
  Signal Processing (ICASSP)}.\hskip 1em plus 0.5em minus 0.4em\relax IEEE,
  2020, pp. 7259--7263.

\bibitem{gutmann2012noise}
M.~U. Gutmann and A.~Hyv{\"a}rinen, ``Noise-contrastive estimation of
  unnormalized statistical models, with applications to natural image
  statistics,'' \emph{Journal of Machine Learning Research}, vol.~13, no. Feb,
  pp. 307--361, 2012.

\bibitem{kolda2009tensor}
T.~G. Kolda and B.~W. Bader, ``Tensor decompositions and applications,''
  \emph{SIAM review}, vol.~51, no.~3, pp. 455--500, 2009.

\bibitem{murphy2012machine}
K.~P. Murphy, \emph{Machine learning: a probabilistic perspective}.\hskip 1em
  plus 0.5em minus 0.4em\relax MIT press, 2012.

\bibitem{gretton2008kernel}
A.~Gretton, K.~Fukumizu, C.~H. Teo, L.~Song, B.~Sch{\"o}lkopf, and A.~J. Smola,
  ``A kernel statistical test of independence,'' in \emph{Advances in neural
  information processing systems}, 2008, pp. 585--592.

\bibitem{read2012goodness}
T.~R. Read and N.~A. Cressie, \emph{Goodness-of-fit statistics for discrete
  multivariate data}.\hskip 1em plus 0.5em minus 0.4em\relax Springer Science
  \& Business Media, 2012.

\bibitem{kankainen1995consistent}
A.~Kankainen, \emph{Consistent testing of total independence based on the
  empirical characteristic function}.\hskip 1em plus 0.5em minus 0.4em\relax
  University of Jyv{\"a}skyl{\"a}, 1995, vol.~29.

\bibitem{feuerverger1993consistent}
A.~Feuerverger, ``A consistent test for bivariate dependence,''
  \emph{International Statistical Review/Revue Internationale de Statistique},
  pp. 419--433, 1993.

\bibitem{panayotov2015librispeech}
V.~Panayotov, G.~Chen, D.~Povey, and S.~Khudanpur, ``Librispeech: an asr corpus
  based on public domain audio books,'' in \emph{2015 IEEE International
  Conference on Acoustics, Speech and Signal Processing (ICASSP)}.\hskip 1em
  plus 0.5em minus 0.4em\relax IEEE, 2015, pp. 5206--5210.

\bibitem{paul1992design}
D.~B. Paul and J.~M. Baker, ``The design for the wall street journal-based csr
  corpus,'' in \emph{Proceedings of the workshop on Speech and Natural
  Language}.\hskip 1em plus 0.5em minus 0.4em\relax Association for
  Computational Linguistics, 1992, pp. 357--362.

\bibitem{peters2018dissecting}
M.~E. Peters, M.~Neumann, L.~Zettlemoyer, and W.-t. Yih, ``Dissecting
  contextual word embeddings: Architecture and representation,'' \emph{arXiv
  preprint arXiv:1808.08949}, 2018.

\end{thebibliography}
